\documentclass[a4paper]{jpconf}
\usepackage{graphicx}
\usepackage[numbers,square,sort&compress]{natbib}
\usepackage{xcolor}
\usepackage{booktabs, array}
\usepackage{listings}
\def \cobj #1{\lstinline[basicstyle=\ttfamily]{#1}}

\usepackage{hyperref}
\usepackage{lineno}

\begin{document}
\title{Hydra: a C++11 framework for data analysis in massively parallel platforms}

\author{A A Alves J\'unior$^1$ and M D Sokoloff$^2$ }

\address{Physics Department - University of Cincinnati\\
400 Geology/Physics Building PO Box 210011
Cincinnati, OH 45221-0011 United States }

\ead{aalvesju@cern.ch$^1$ and sokoloff@ucmail.uc.edu$^2$ }

\begin{abstract}
Hydra is a header-only, templated and C++11-compliant framework designed to perform the typical bottleneck  
calculations found in common HEP data analyses on massively parallel platforms.
The framework is implemented on top of the C++11 Standard Library and a variadic version of the Thrust library
and is designed to run on Linux systems, using OpenMP, CUDA and TBB enabled
devices. This contribution summarizes the main features of Hydra. A basic description
of the overall design, functionality and user interface is provided, along with some code examples and measurements
of performance.
\end{abstract}

\section{Introduction}
\vspace{0.2cm}
Despite the ongoing efforts of modernization, a large fraction of the software used in High Energy Physics (HEP) is legacy. 
It mostly consists of libraries assembling single threaded,
Fortran and C++03 mono-platform routines. Concomitantly, HEP experiments keep collecting samples with unprecedented large statistics,  while
data analyses become increasingly complex. Even deploying powerful computers, it is common 
to spend days performing calculations to reach a result, and they often need re-tuning anyway. 

On the other hand, computer processors will not increase clock frequency anymore in order to reach higher performance. Indeed, the current road-map to improve overall
performance is to deploy different levels of concurrency. One effect of which is the proliferation of multi-thread friendly and multi-platform environments 
among HPC data centers. Unfortunately, HEP software is not completely prepared yet to fully exploit concurrency and to deploy more opportunistic computing strategies.

Hydra proposes a computing model to approach these issues. The Hydra framework provides collection of
parallelized high-level algorithms, addressing some of of typical computing bottlenecks commonly found in HEP,
and a set of optimized containers and types, through a modern and functional interface, allowing to enhance HEP software productivity and
performance, keeping the portability between NVidia GPUs, multi-core CPUs and other devices compatible with CUDA, TBB and OpenMP computing models.

\section{Design highlights}
\vspace{0.2cm}
Hydra is a C++11 template framework organized using a variety of static polymorphism idioms and patterns. This ensures the predictability of the stack size at compile time, 
which is critical for stability and performance when running on GPUs and minimizes the overhead introduced by the user interface
when engaging the actual calculations. Furthermore, the combination of static polymorphism and templates
allows exposure of the maximum amount of code to the compiler, in the context in which the code will be used, contributing to activate many 
compile time optimizations that could not be accessible otherwise. Hydra's interface and implementation details extensively deploy patterns and idioms
that enforce thread-safety and efficient memory access and management. The following list summarizes some of the main design choices adopted in Hydra:

\begin{itemize}
 \item Hydra provides a set of optimized STL-like containers that can store multidimensional data-sets using 
 SoA\footnote{Structure of arrays or SoA is a layout separating elements of a structure
 into one parallel array per field. This ease the data manipulation with SIMD instructions and, if only a specific field of the structure is needed, 
 only this field can to be iterated over, allowing more data to fit onto a single cache line.}
 \item Data handled using iterators and all classes manage allocated resources using RAII idiom.
 \item The framework is type and thread-safe.
 \item No limitation on the maximum number of dimensions that containers and algorithms can handle.  
\end{itemize}

The types of devices in which Hydra can be deployed are classified by back-end type, according to the device compatibility with certain computing models.
Currently, Hydra supports four back-ends, which are CPP, OpenMP, CUDA and TBB. Code can be dispached and executed in all supported back-ends concurrently and asynchronously 
in the same program, using the suitable policies represented by the symbols \cobj{hydra::omp::sys}, \cobj{hydra::cuda::sys}, \cobj{hydra::tbb::sys}, \cobj{hydra::cpp::sys}, \cobj{hydra::host::sys}
and \cobj{hydra::device::sys}. These policies define the memory space where resources should be allocated to run algorithms and store data.

For mono-back-end applications, source files written using Hydra and standard C++ compile for GPU and CPU just
exchanging the extension from .cu to .cpp and one or two compiler flags. There is no need for re-factory code.

\section{Basic features} 
\label{sec:features}
\vspace{0.2cm}
Currently, Hydra provides a collection of
parallelized high-level algorithms, addressing some computing-intensive tasks commonly found in data analyses in HEP.
The available high-level algorithms are listed below:

\begin{itemize}
 \item Multidimensional p.d.f. sampling.
 \item Parallel function evaluation on multidimensional data-sets.
 \item Five fully parallelized numerical integration algorithms: Genz-Malik, self-adaptive and static Gauss-Kronrod quadratures,
 plain, self-adaptive importance sampling and phase-space Monte Carlo integration.  
 \item Phase-space Monte Carlo generation, integration and modeling. 
 \item Interface to \cobj{ROOT::Minuit2}\cite{Minuit} minimization package, allowing to accelerate maximum likelihood fits over multidimensional large data-sets.   
 \item Parallel implementation of the S-Plots\cite{Pivk:2004ty} technique prescription, for statistical unfolding data distributions.
\end{itemize}

Hydra also provides two multidimensional containers optimized to store large POD data-sets with dimensions represented by numbers with the same or different types. 
A significant number of facilities are also available to filter, copy and perform other operations. In the following subsections, some of the basic features are 
described. 

\subsection{Functors and arithmetic operators}
\vspace{0.2cm}
In all situations where custom calculations need to be performed,
Hydra framework instantiates algorithms to call the user's code that is passed using functors or C++11 lambdas.
Hydra adds features and type information to generic functors using the CRTP idiom. To be compliant with Hydra's interface,
the functors need to derive from the \cobj{hydra::BaseFunctor<Functor,ReturnType,NParameters>} class and to implement
the \cobj{ __host__ __device__ Evaluate(...)} method. Functor's values can be cached. 
The \autoref{lst:functor} shows how to implement a simple functor to calculate 
a Gaussian function.

\begin{lstlisting}[caption={ Example of implementation of a functor representing a Gaussian function in Hydra.
  The access to the Hydra interface and type information is implemented deriving the functor 
  from the \cobj{hydra::BaseFunctor<Functor, ReturnType, NParameters>} base class. The functor's parameters, 
  in this case the mean and sigma of the Gaussian, are accessible in the scope of the functor via the symbols \cobj{_par[i]}
  or in the functor instantiation scope via accessors \cobj{SetParameter(...)} and \cobj{GetParameter(...)}.},
  label=lst:functor, float, basicstyle=\small,linewidth=.90\textwidth]
  
struct Gaussian: public hydra::BaseFunctor<Gaussian,double,2>
{
  //implementing the Evaluate() method of the Gaussian functor.
  //unsigned int n : number of arguments
  //double *x: pointer to the array of arguments
   __host__ __device__
  inline double Evaluate(unsigned int n, double* x){
    double _m2 = x[0]*_par[0]*_par[0];
    double _s2 = _par[1]*_par[2];
    
    return  exp(-m2/(2.0 * _s2 ))/( sqrt(2.0*_s2*pi));
  }
};
\end{lstlisting}

%
% ...
 
\subsection{C++11 Lambdas}
 \vspace{0.2cm}
The user can define a C++11 lambda function and convert it into a Hydra functor using the function 
\cobj{hydra::wrap_lambda()}. The wrapped lambda will have access to all functionality of Hydra. The \autoref{lst:lambda1} shows 
a basic example of usage for this functionality.
Hydra also supports the usage of C++11 lambdas as fit models. In this case, the lambda function 
needs to have a different signature and the list of fit parameters needs to be passed to \cobj{hydra::wrap_lambda()}. 

\begin{lstlisting}[caption={Example of usage of \cobj{hydra::wrap_lambda()}
to wrap a  simple lambda function multiplying 
sin(x) by two.}, label=lst:lambda1, float, basicstyle=\small,linewidth=.90\textwidth]
...
double two = 2.0;
//define a simple lambda and capture "two"
auto my_lambda = [=] __host__ __device__ (unsigned int n, double* x){
         return two*sin(x[0]); 
     };
//convert it into a Hydra functor
auto my_lambda_wrapped  = hydra::wrap_lambda(my_lambda);
...
\end{lstlisting}

% ...

\subsection{Arithmetic operators}
\vspace{0.2cm}
Hydra overloads the basic arithmetic operators for all objects deriving from  \cobj{hydra::BaseFunctor}. 
Composition of functors is supported as well.
For example, the lines shown in \autoref{lst:operators} are completely legal C++11 code.

\begin{lstlisting}[caption={Using overloaded arithmetic operators to build up expressions using functors. 
 If \cobj{A}, \cobj{B} and \cobj{C} are Hydra functors, the code above is completely legal.},
label=lst:operators, float, basicstyle=\small,linewidth=.90\textwidth]
 ...
 //basic arithmetic operations
 auto A_plus_B  = A + B;
 auto A_minus_B = A - B;
 auto A_times_B = A * B;
 auto A_per_B   = A/B; 
 //any composition of basic operations
 auto any_functor = (A - B)*(A + B)*(A/C);
 // C(A,B) is represented by:
 auto compose_functor = hydra::compose(C, A, B)
 ...
\end{lstlisting}


\subsection{Containers}
\vspace{0.2cm}
Hydra framework provides one dimensional STL-like vector container for each supported back-end, aliasing the underlying Thrust 
types. Beyond this, Hydra implements two native multidimensional containers: \cobj{hydra::multivector} and   \cobj{hydra::multiarray}.
In these containers, the data corresponding to each dimension is stored in contiguous memory regions and when the container is traversed, each entry is accessed as 
a \cobj{hydra::tuple}, where each field holds a value corresponding to a dimension. Both classes implement an interface completely compliant with a STL vector and
also provides constant and non-constant accessors for the single dimensional data. The container \cobj{hydra::multivector} is suitable to store data-sets where the dimensions are represented by entries with different POD types.   
\cobj{hydra::multiarray} is designed to store data-sets where all dimensions are represented by the same type. Data is copiable across different back-ends.

\section{Code examples and performance measurements}
\label{sec:examples}
\vspace{0.2cm}
\subsection{Multidimensional numerical integration}
\begin{figure}[t]
\begin{center}
  \includegraphics[width=.7\linewidth]{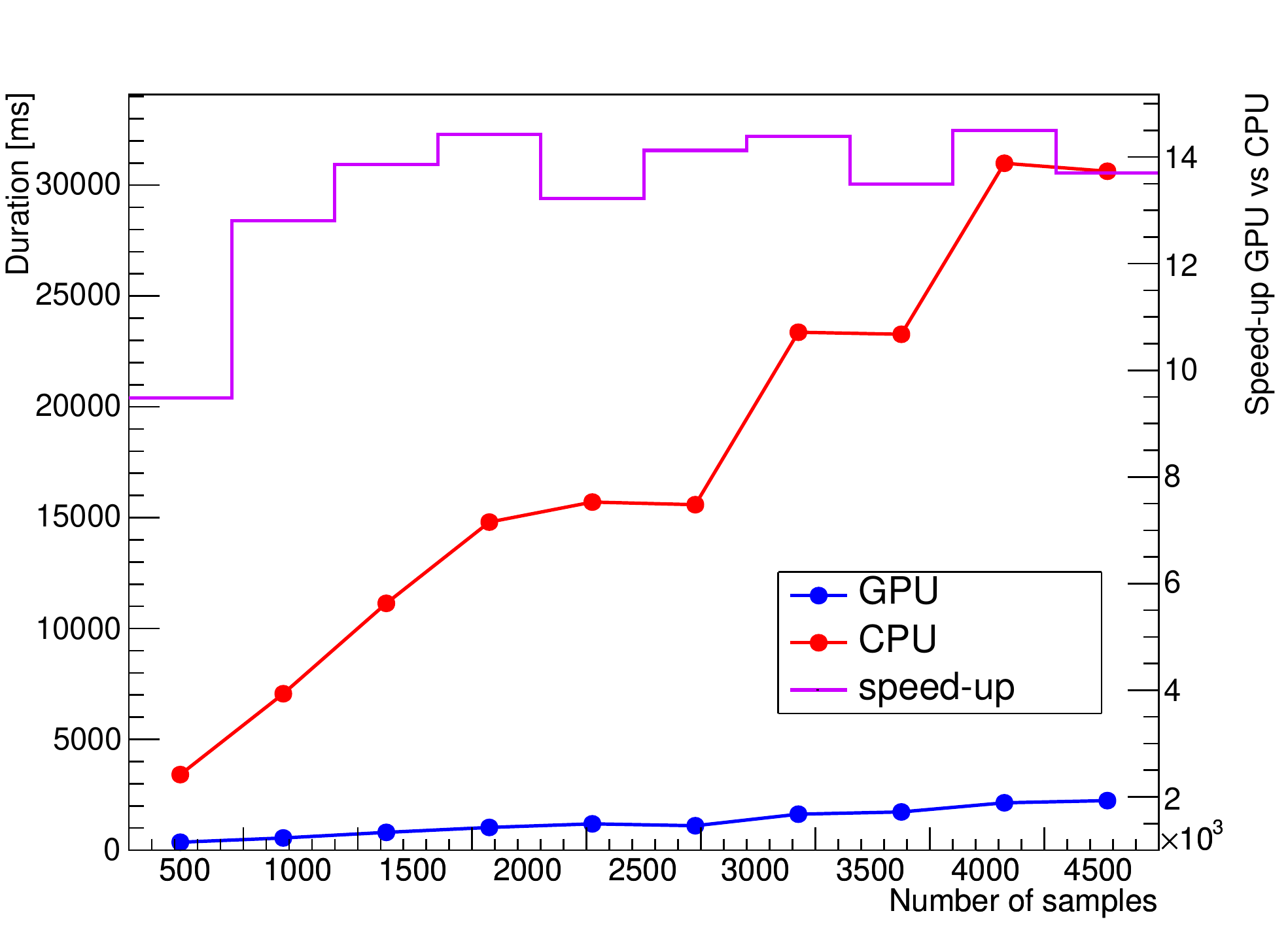}
\end{center}
 \caption{Performance of Hydra's VEGAS implementation as a function of the number of integrand calls per iteration.
 The integrand is a 10-dimensional Gaussian distribution. The GPU model is Tesla K40c
 and the CPU is Intel Xeon(R) CPU E5-2680 v3 @ 2.50GHz (one thread).  }
 \label{fig:vegas}
\end{figure}

\vspace{0.2cm}
The VEGAS algorithm\cite{PETERLEPAGE1978192} samples the integrand and adapts
itself, so that the sampling points are concentrated in the regions that make the largest contribution to
the integral. Hydra's implementation follows the structure of the corresponding GSL algorithm, but parallelizes
the integrand evaluations and accumulations in each iteration. There is no limit in the number of dimensions the integrator can handle.
\autoref{fig:vegas} shows the performance of Hydra's VEGAS implementation. 
 
\subsection{Phase-Space Monte Carlo}
\vspace{0.2cm}
Phase-Space Monte Carlo simulates the kinematics of a particle with a given four-momentum
decaying to a n-particle final state, without intermediate resonances. Samples 
of phase-space Monte Carlo events are widely used in HEP studies where 
the calculation of phase-space volume is required as well as 
a starting point to implement and to describe the properties of models with one or more resonances or
to simulate the response of the detector to decay's products\cite{James:275743}. 
Hydra provides an implementation of the Raubold-Lynch method\cite{James:275743}
and can generate the full kinematics of decays with any number of particles in the final state.
Sequential decays, evaluation of model, production of weighted and unweighted samples and many other features are 
also supported. \autoref{lst:phsp} shows how to use the  phase-space Monte Carlo generator to produce a sample 
with \cobj{ndecays} three-body decays. \autoref{fig:phsp} shows the phase-space generator performance 
as a function of the sample size.

\begin{lstlisting}[caption={This snippet shows how to use the Hydra's phase-space Monte Carlo generator to
produce a sample with \cobj{ndecays} three-body decays.},label=lst:phsp, float, basicstyle=\small,linewidth=.90\textwidth]   
//Masses of the particles
hydra::Vector4R Mother(mother_mass, 0.0, 0.0, 0.0);
double Daughter_Masses[3]{daughter1_mass, daughter2_mass, daughter3_mass };
//Create PhaseSpace object
hydra::PhaseSpace<3> phsp(Mother_mass, Daughter_Masses);
//Allocate the container for the events 
hydra::Events<3, device> events(ndecays);
//Generate 
phsp.Generate(Mother, events.begin(), events.end());
\end{lstlisting}


\begin{figure}[t]
\begin{center}
  \includegraphics[width=.7\linewidth]{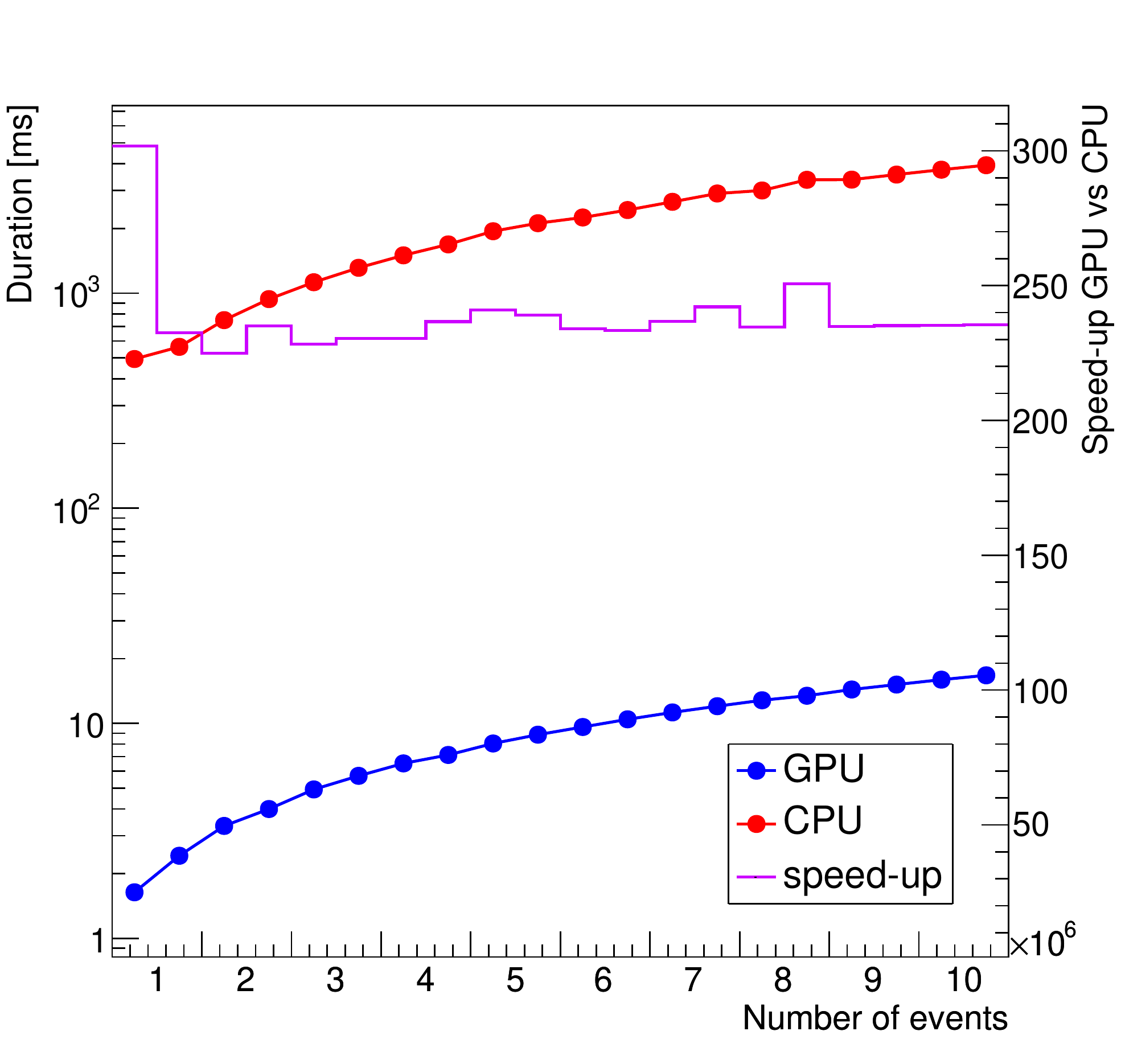}
\end{center}
\caption{The performance of the phase-space generator
as a function of the sample size. The GPU model is Tesla K40c
 and the CPU is Intel Xeon(R) CPU E5-2680 v3 @ 2.50 GHz (one thread). }
 \label{fig:phsp}
 \end{figure}

\subsection{Interface to Minuit2}
\vspace{0.2cm}
Hydra implements an interface to \cobj{ROOT::Minuit2} that parallelizes the FCN calculation \cite{Minuit}.
This dramatically accelerates the calculation over large data-sets. Hydra normalizes the pdfs 
on-the-fly using analytical or numerical integration algorithms provided by the framework and handles data using iterators. 

\autoref{lst:minuit} shows how to build a simple model composed of two pdfs, respectively
representing a Gaussian and an exponential distributions. The model is used to perform an extended likelihood fit.
The FCN object provided by Hydra parallelizes the function calls in the back-end the data is stored in. 
The processing of a sample with 20 million events takes 4.8 seconds in a Tesla K40c and 299.8 seconds in a
Intel Xeon(R) CPU E5-2680 v3 @ 2.50 GHz (one thread), resulting in a speed-up of a factor $\sim$62x.

\begin{center}
\begin{lstlisting}[caption={This code snippet shows how to build simple model composed by two pdfs, respectively representing a Gaussian and an exponential distributions.
    The model is used to perform a extended likelihood fit, able to predict the yields of each component. The FCN object provided by Hydra, 
    will parallelize the function calls in the back-end the data is stored. }, label=lst:minuit, float, basicstyle=\small,linewidth=.90\textwidth]
//analytical integral functors 
GaussAnalyticIntegral GaussIntegral(min, max);
ExpAnalyticIntegral   ExpIntegral(min, max);
//build the pdfs
auto Gaussian_PDF    =  hydra::make_pdf(Gaussian, GaussIntegral);
auto Exponential_PDF =  hydra::make_pdf(Exponential, ExpIntegral);

//make a extended pdf model
std::array<hydra::Parameter*, 2> yields{NGaussian, NExponential};
auto Model = hydra::add_pdfs(yields, Gaussian_PDF, Exponentia_PDF );
//get the FCN
auto Model_FCN = hydra::make_loglikehood_fcn(Model, data_d.begin(), data_d.end());
//pass the FCN to Minuit2
...
\end{lstlisting}
\end{center}

%
%
% ...

\begin{figure}[t]
\begin{center}
\includegraphics[width=.6\linewidth]{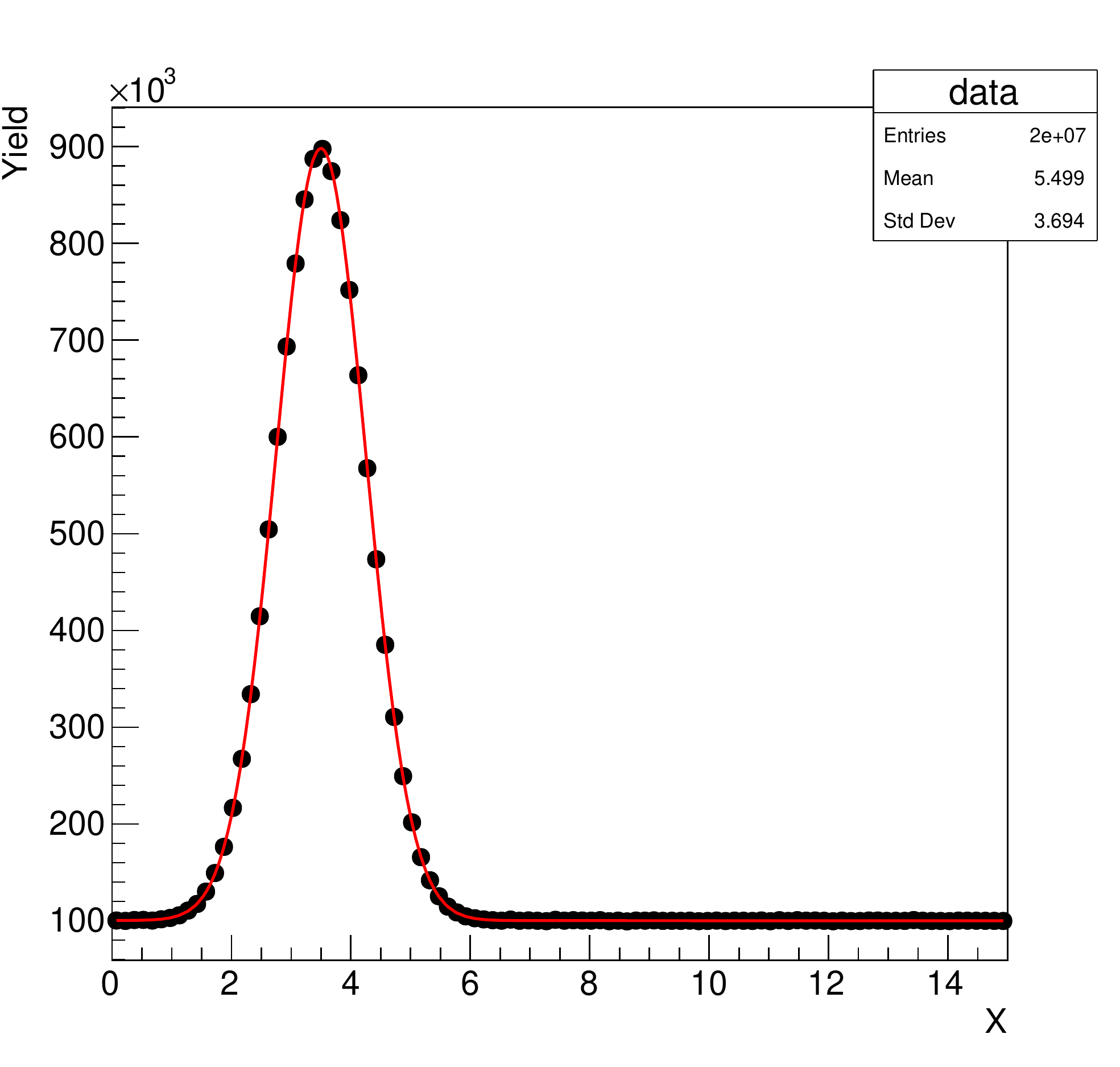}
\end{center}
\caption{Result of the extended unbined likelihood fit performed over a 20 million events sample. }
 \label{fig:phsp}
 \end{figure}
 
\section{Summary}
\vspace{0.2cm}
The basic design, performance and functionality of the header-only, C++11-compliant framework Hydra
have been introduced. Some of the basic interfaces and algorithms are discussed in \autoref{sec:features}. 
The performance measurements for running some of Hydra's algorithms using CUDA and one CPU thread are discussed in \autoref{sec:examples}, and show that Hydra can be 
up to 250 times faster than conventional software, depending on the graphics card. For CPU multi-threads back-ends, TBB and OpenMP, if the problem size is large and calculations
pay the cost of thread creation and destruction, the algorithms scale linearly with the number of threads deployed. 
Since Hydra is header only, no additional building process needs to be done beyond the inclusion of the required
headers. Hydra has been presented at several conferences, including NVidia's GTC 2017\cite{GTC}. It is  currently being used
in a data analysis aiming to measure the charged Kaon mass at the LHCb Experiment at CERN\cite{kaon-mass}. 
The initial development of Python bindings for Hydra was one of the projects of Google Summer of Code 2017\cite{GSoC}.

Hydra is open source and released under GPL version 3 license. 
The project is hosted on GitHub at \url{https://github.com/MultithreadCorner/Hydra}.

\section{Acknowledgments}
\vspace{0.2cm}
This work was performed with support from NSF Award PHY-1414736. 
NVidia  provided K40 GPUs for our use through its University Partnership program.

\bibliography{references}

\end{document}